\documentclass[10pt,letterpaper]{article}
\usepackage[top=0.85in,left=2.75in,footskip=0.75in,marginparwidth=2in]{geometry}

\usepackage[utf8]{inputenc}

\usepackage{cite}

\usepackage{nameref,hyperref}
\usepackage{amsmath, amsthm, amssymb, amsfonts}
\usepackage{footnote}

\usepackage[right]{lineno}

\usepackage{microtype}
\DisableLigatures[f]{encoding = *, family = * }

\raggedright
\usepackage{changepage}

\usepackage[aboveskip=1pt,labelfont=bf,labelsep=period,singlelinecheck=off]{caption}

\makeatletter
\renewcommand{\@biblabel}[1]{\quad#1.}
\makeatother

\usepackage{lastpage,fancyhdr,graphicx}
\usepackage{epstopdf}
\fancyheadoffset[L]{2.25in}
\fancyfootoffset[L]{2.25in}

\usepackage{color}

\definecolor{Gray}{gray}{.25}

\usepackage{graphicx}

\usepackage{sidecap}

\usepackage{wrapfig}
\usepackage[pscoord]{eso-pic}
\usepackage[fulladjust]{marginnote}
\reversemarginpar

\begin{document}
\vspace*{0.35in}

% title goes here:
\begin{flushleft}
{\Large
\textbf\newline{Generalized Autoregressive Neural Network Models.}
}
\newline
% authors go here:
\\
Renato Rodrigues Silva\textsuperscript{1,*}
\\
\bigskip
\bf{1} Federal University of Goias
\\
\bigskip
* renato.rrsilva@ufg.br
\end{flushleft}

\section*{Abstract}
A time series is a sequence of observations taken sequentially in time \cite{boxjen76, bookHamilton1994, bookwei2006}. The autoregressive integrated moving average is a class of the model more used for time series data.However, this class of model has two critical limitations. It fits well only Gaussian data with the linear structure of correlation. Here, I present a new model named as generalized autoregressive neural networks, GARNN. The GARNN is an extension of the generalized linear model where the mean marginal depends on the lagged values via the inclusion of the neural network in the link function. A practical application of the model is shown using a well-known poliomyelitis case number, originally analyzed by \cite{paperZegerandQaqish1988}.

\section*{Introduction}

A time series is a sequence of observations taken sequentially in time \cite{boxjen76, bookHamilton1994, bookwei2006}. The model more widely used for time series analysis is the autoregressive integrated moving average, ARIMA. One of the reasons for this popularity is due to the well known Box-Jenkins methodology \cite{boxjen76}. 

In summary, the Box-Jenkins method consists of three steps that are carried out repeatedly until to find a suitable model. The first stage is identification of the model. The identification is done by use of the autocorrelation (ACF) and partial autocorrelation (PACF) functions. The second stage is the estimation, usually done by maximizing the likelihood function. The last stage consists of plotting autocorrelation and partial autocorrelation of the residuals \cite{boxjen76, bookHamilton1994, bookwei2006}. 

ARIMA models have proven over time to be an efficient option to make forecasting. However, this class of model has two critical limitations. The ARIMA model is adequate for only Gaussian data. Moreover, it assumes a linear structure of the correlation, such as the autoregressive process and or moving average process \cite{paperZhang2003}.

There a vast literature about non-Gaussian time series models. For example, \cite{paperBenjamin2003} proposed a class of generalized autoregressive moving average (GARMA) models. They developed a method of estimation of the parameters based on iteratively reweighted least squares algorithm and studied stationary properties of the model. \cite{paperCordeiroAndrade2009} introduced a new class of transformed generalized linear models and used these models to time series data to extend the GARMA models discussed by \cite{paperBenjamin2003}. \cite{paperFonseca2018} proposed a generalized autoregressive  model based on a bimodal Birnbaum–Saunders distribution. They presented and discussed parameter estimation, hypothesis testing inference, residual analysis and develop prediction intervals. \cite{paperBayer2018} beta seasonal autoregressive moving average for modelling time series data, which assumes values in the unit interval.

Although the autoregressive and or moving average process is a reasonable assumption for time series, the use of the more complex structure of correlations sometimes should be used. An option to handle this is the use of a time lag neural network \cite{paperZhang2003}. \cite{paperLapedesFarber1987} were the first authors to apply successfully artificial neural networks to make forecasting in the time series data. \cite{paperParkSanderberg1991} reported that artificial neural networks outperformed regression models in forecasting total load consumption. \cite{summarySen2019} developed a forecasting model that combines a global matrix factorization and a temporal convolutional neural network. 

However, none of the models described up to now, handle with the two mentioned limitations of the ARIMA models in the same time. Here, I present a new model named as generalized autoregressive neural networks, GARNN. This model deal with non-Gaussian time series with a linear or non-linear structure correlation. I extended the generalized linear model adding a time lag neural network in the link function. In the next sections I introduce the details about the definition of the GARNN model and submodels, parameters estimation, model selection, predictions. Subsequently, I present a pratical application with real data with discussion.

\section*{Statistical Model - GARNN}

The GARNN model contains four components: random component, systematic component, link function and time lagged neural networks  component. The random component denoted by $Y_{t}|{\cal F}_{t-1}$ follows a distribution which belongs to exponential family

\begin{equation} \nonumber
  f(y_{t}|{\cal F}_{t-1}) = \exp\left\{  \frac{y_t\alpha_t - b(\alpha_t)}{\varphi} + d(y_t,\alpha_t)\right\} ,
\end{equation}
where $E[Y_{t}|{\cal F}_{t-1}] =  b^{'}(\alpha_t) = \mu_t$ e $Var[Y_{t}|{\cal F}_{t-1}] = \varphi b^{''}(\alpha_t) = \varphi V_t$;
$t = 1, \ldots, n$ and $y_1, \ldots, y_n$ are realizations of the response variables.

The systematic component is comprised of the regression coefficients and covariables. The covariables  could be variables that capture cycles and or seasonalities, such as $\cos\left(\frac{2\pi t}{12}\right)$ and $\sin\left(\frac{2\pi t}{12}\right)$, or, some explanatory variables related to $y_t$, as for example, calendar effects. The link function $g(.)$ is a monotonic and differentiable function that describes the relationship between the expected value of $Y_{t}|{\cal F}_{t-1}$, denoted by $\mu_t$, and the non linear predictor $\eta_t$. In turn, $\eta_t$ is comprised of the linear predictor of the systematic component, $\mathbf{x}^{'}_{t}\boldsymbol{\beta}$, added to the time lagged neural networks. The time-lagged neural networks component is a single layer feedforward neural network composed by inputs that are lagged values of the series and lagged values of the systematic components. This neural net aims to model the dependency of values observed \cite{paperLapedesFarber1987}. The expression of the $g(\mu_t)$ is given by

\begin{equation} \nonumber
g(\mu_t) = \eta_t = \mathbf{x}^{'}_{t}\boldsymbol{\beta} + \sum_{i=1}^{I}\rho_i h(G_{it})
\end{equation}
where $h(.)$ is the function activation of the time-lagged neural network, $I$ is the number of nodes, $\rho_i$ is the i-th weight of output layer, $G_{it} = \sum_{j=i}^p \omega_{ij} z_{t-j}$ is the output layer linear predictor of the neural net and $\omega_{ij}$ is the weight of the input defined at i-th node and j-th lag, $z_{t-j} = \frac{y_{t-j} - \bar{y}}{s_y}$ is the standardized input of the i-th node and j-th lag, 
$\bar{y}$ and $s_y$ is the sample mean and standard desviation of the $y_1, \ldots, y_n$, respectively. Following \cite{paperTrapletti1970}, only bounded activation functions were considered.

\subsection*{Particular cases}

\subsubsection*{GARNN Models for Counting, Binary and Proportion Data }

The probability density function of Poisson GARNN model is 

\begin{equation}
f(y_{t}|{\cal F}_{t-1}) = \exp\left\{y_t \log(\mu_t) - \mu_t \log(y_t!)\right\},
\end{equation}
with canonical link function given by $g(\mu_t) = \log(\mu_t).$

The probability density function of the Binomial GARNN model is the following

\begin{equation}
f(y_{t}|{\cal F}_{t-1})  = \exp\left\{y_t \log\left( \frac{\mu_t}{m-\mu_t}\right) +  m\log\left( \frac{m - \mu_t}{m}\right) + \log{ m \choose y_t }\right\}.
\end{equation}

In this case, the canonical link function is called logistic function, defined by $g(\mu_t) = \frac{1}{1+e^{-\eta_t}}.$

Assuming $k$ known, the probability density function of the Negative Binomial GARNN model given by:

\begin{equation} \nonumber
f(y_{t}|{\cal F}_{t-1}) = \exp\left\{ k \log\left(\frac{k}{\mu_t + k}\right) + y_t \log\left(\frac{\mu_t}{\mu_t + k}\right) + \log\left(\frac{\Gamma(k+y_t)}{\Gamma(y_t + k)\Gamma(k)}\right)  \right\},
\end{equation}
where the link function used is logarithmic $\log\left(\frac{k}{\mu_t + k}\right).$

\subsubsection*{GARNN Models for Continuous Data}

Suppose that $Y_{t}|{\cal F}_{t-1} \sim N(\mu_t,\sigma^2)$ the probability density function of the Normal GARNN model is

\begin{equation} \nonumber
f(y_{t}|{\cal F}_{t-1}) = \frac{1}{\sqrt{2\pi\sigma^2}} \exp\left\{ -\frac{1}{2}\left[\frac{(y_t - \mu_t)^2}{\sigma^2}\right] \right\},
\end{equation}
with identity canonical link function.

The Gamma GARNN model is used for modelling assymetric data. The probability density function of the is given by

\begin{equation} \nonumber
f(y_{t}|{\cal F}_{t-1}) = \exp\left\{\nu\left[y_t\left(-\frac{1}{\mu_t}\right) - \log \mu_t\right] \nu \log \nu y_t - \log y_t - \log \Gamma(\nu)\right\},
\end{equation}
with reciprocal canonical link function $g(\mu_t) = \frac{1}{\eta_t}$.

\section*{Estimation}

Let's consider $\boldsymbol{\theta}^{'} = \left(\boldsymbol{\beta}^{'}, \boldsymbol{\omega}^{'}, \boldsymbol{\rho}^{'} \right)$ the parameters to be estimated. Following \cite{paperBenjamin2003}, the log likelihood of the data $\left\{y_{m+1}, \ldots, y_n\right\}$ conditional on the first $m \geq p$ observations $\left\{y_{1}, \ldots, y_m\right\}$  is defined by

\begin{equation} \nonumber
l(\boldsymbol{\theta}) = \sum_{t=m+1}^n l_t(\alpha_t, \varphi) = \sum_{t=m+1}^n \frac{y_t\alpha_t -b(\alpha_t)}{\varphi} + \sum_{t=m+1}^n d(y_t, \varphi).
\end{equation}

The score function is defined using chain rule for derivatives, i.e,

\begin{align}
U(\boldsymbol{\theta}) = \frac{\partial l(\boldsymbol{\theta})}{\partial\boldsymbol{\theta}}, \label{eqScore}
\end{align}
where

\begin{align}
U(\theta_s) = \frac{\partial l\theta_s}{\partial\theta_s} =  \frac{1}{\varphi}\sum_{t=m+1}^n
\frac{d l_t(\alpha_t \varphi)}{d \alpha_t} \frac{d \alpha_t}{d \mu_t} \frac{d \mu_t}{d \eta_t}\frac{\partial \eta_t}{\partial \theta_s} = \sum_{t=m+1}^n \frac{d \mu_t}{d \eta_t}\frac{\partial \eta_t}{\partial \theta_s} \frac{(y_t - \mu_t)}{V_t}. \label{eq1}
\end{align}

Using matrix notation the expression given by (\ref{eqScore}) can rewritten as follows

\begin{align} \nonumber
U(\boldsymbol{\theta}) 
=&
\left(\frac{\partial \boldsymbol{\eta}}{\partial\boldsymbol{\theta}^{'}}\right)^{'}\mathbf{W}(\mathbf{y}-\boldsymbol{\mu}) \\
=& 
\left[
\begin{matrix}
\frac{\partial \eta_{m+1}}{\partial\theta_1} & \ldots &
\frac{\partial \eta_{n}}{\partial\theta_1} \\
\vdots \\
\frac{\partial \eta_{m+1}}{\partial\theta_{S+I(p+1)}} & \ldots &
\frac{\partial \eta_{n}}{\partial\theta_{S+I(p+1)}} 
\end{matrix} 
\right]
\left[
\begin{matrix}
\frac{1}{V_{m+1}  \frac{d \eta_{m+1}}{d \mu_{m+1}}  } & \ldots & 0 \\  
& \ddots & \\
0 & \ldots & \frac{1}{V_{n}  \frac{d \eta_{n}}{d\mu_n}  } 
\end{matrix} 
\right]
\left[
\begin{matrix}
y_{m+1} - \mu_{m+1} \\
\vdots \\
y_{n} - \mu_{n}
\end{matrix} 
\right], \nonumber
\end{align}
where $S$ is the number of the covariables and $V_t$ is the variance function.

Provided also that the link function $g(.)$ is canonical \cite{bookmccullagh1989}, the score function becomes to

\begin{equation} \nonumber
U(\boldsymbol{\theta}) = \left(\frac{\partial \boldsymbol{\eta}}{\partial\boldsymbol{\theta}^{'}}\right)^{'}(\mathbf{y}-\boldsymbol{\mu}).
\end{equation}

where $\left(\frac{\partial \boldsymbol{\eta}}{\partial\boldsymbol{\theta}^{'}}\right)$ is defined by 

\begin{align}
\frac{\partial \eta_t}{\partial \rho_i} =& h(G_{it})  \nonumber \\
\frac{\partial \eta_t}{\partial \omega_{ij} } =& \rho_i \frac{d h}{d G_{it}}z_{t-j} . \nonumber \\
\frac{\partial \eta_t}{\partial \beta_s } =& x_{ts}, \nonumber \\
\end{align}
or alternatively, by its vectorized version 
\begin{align}
\frac{\partial \eta_t}{\partial \boldsymbol{\rho}^{'}} =& \left( h(G_{1t}), \ldots, h(G_{It})   \right)^{'}, \label{back1} \\
\frac{\partial \eta_t}{\partial \boldsymbol{\omega}^{'}} =& \left(\boldsymbol{\rho} \circ \frac{dh}{d\mathbf{G}_t}\right) \circ \left( z_{t-1}, \ldots,  z_{t-p} \right), \label{back2} \\
\frac{\partial \eta_t}{\partial \boldsymbol{\beta}^{'} } =& \mathbf{x}_t^{'}  \label{back3}
\end{align}
where $\boldsymbol{\rho} = \left( \rho_1, \ldots, \rho_I \right)^{'}$  and $\frac{dh}{d\mathbf{G}_t} = \left(\frac{dh}{dG_{1t}}, \ldots, \frac{dh}{dG_{It}} \right)^{'}$.

The score function does not have an analytical solution, hence the Broyden–Fletcher–Goldfarb–Shanno (BFGS) algorithm  \cite{bookNocedal2006} was applied. 

The dispersion parameter is unknown for both Normal, Gamma and Negative Binomial GARNN models. In that case, the vector parameters the score function is defined by

\begin{equation} \nonumber
U(\boldsymbol{\theta},\varphi) = -\sum_{t=m+1}^n \frac{y_t\alpha_t -b(\alpha_t)}{\varphi^2} + \sum_{t=m+1}^n \frac{\partial d(y_t, \varphi)}{\partial \varphi}
\end{equation}

For Normal GARNN model, the score function relative to $\sigma^2$ parameter is given by

\begin{equation}
U_{\sigma^2}(\boldsymbol{\theta},\sigma^2) \propto -\frac{n-(m+1)}{2 \sigma^2} -\frac{1}{2}\sum_{t=m+1}^n\left[\frac{(y_t - \mu_t)^2}{(\sigma^2)^2}\right]
\end{equation}
and the maximum likelihood estimator for $\sigma^2$ is given by

\begin{equation} \nonumber
\hat{\sigma}^2 = \frac{1}{2 (n-(m+1))}\sum_{t=m+1}^n(y_t - \hat{\mu}_t)^2.
\end{equation}

\cite{paperCordeiroMcCullagh1991} suggested that the dispersion parameters of the Gamma GARNN model can be estimated from solution of the non linear equation $\log(\hat{\nu}^{-1}) - \frac{d\Gamma(\hat{\nu}^{-1})}{d\hat{\nu}^{-1}} = \frac{2D_p}{n-(m+1)}$, where $D_p = 2\sum_{t=m+1}^n\left[\log(\frac{\hat{\mu}_t}{y_t}) + \frac{y_t - \hat{\mu}_t}{\hat{\mu}_t}\right]$ . The $k$ of the Negative Binomial GARNN can be estimated via profile likelihood \cite{paperZhang2003}.

\section*{Model Selection}

Comparison between two nested models can be done using analysis of Deviance and or F-test \cite{paperBenjamin2003, bookmccullagh1989}.

Let's suppose two nested models with total number of parameters $\kappa_0$ and $\kappa_1$ and log likelihoods $\hat{l}_0$ and $\hat{l}_1$. If dispersion parameters is known, under null hypothesis,  the statistics $\Lambda = \frac{(\hat{D}_0 - \hat{D}_1)}{\varphi}$ follows a Chi-squared distribution with $(\kappa_1-\kappa_0)$ degree of freedom, where
$\frac{\hat{D}_0}{\varphi} = -2 \hat{l}_0$ and $\frac{\hat{D}_1}{\varphi} = -2 \hat{l}_1.$ Otherwise, under null hypothesis, the statistics $\lambda_F = \frac{(\hat{D}_0 - \hat{D}_1)}{\hat{\varphi}(\kappa_1-\kappa_0)}$ follows approximately F distribution with $(\kappa_1 - \kappa_0)$, $(n - m - \kappa_1)$ degrees of freedom. For non nested models, model selection can be done via Akaike Information Criterion \cite{paperAkaike1974}.

\section*{Prediction}

One of the main goals in the time series analysis is the prediction and 
forecasting ahead time. Prediction can be expressed mathematically as

\begin{align}
\hat{y}_{t} = \hat{\mu}_{t} = g^{-1}(\hat{\eta}_{t}), \nonumber
\end{align}
for $t= m+1, \ldots, n$, where

\begin{align}
\hat{\eta}_{t} = \eta_t = \mathbf{x}^{'}_{t}\boldsymbol{\beta} + \sum_{i=1}^{I}\rho_i \sum_{j=i}^p 
\hat{\omega}_{ij} z_{t-j} . \nonumber
\end{align}

Considering $n$ the forecasting origin and $r >0$, the $n+r$-th step ahead forecast is done recursively.
For instance, assuming $r = 1$, the forecast equation can be expressed as

\begin{equation}
\hat{\mu}_{n+1} = g^{-1}\left(\mathbf{x}^{'}_{n+1}\boldsymbol{\hat{\beta}} + \sum_{i=1}^{I}\hat{\rho}_i h\left( \sum_{j=i}^p 
\hat{\omega}_{ij} \left(\frac{y_{n} - \bar{y}}{s_y} \right)  \right) \right),
\end{equation}
In that case, the $\mathbf{x}^{'}_{n+1}$ should have already been predicted.

Generalizing, the $n+r$-th step ahead forecast is defined via 

\begin{equation}
\hat{\mu}_{n+r} = g^{-1}\left(\mathbf{x}^{'}_{n+r}\boldsymbol{\hat{\beta}} + \sum_{i=1}^{I}\hat{\rho}_i h\left( \sum_{j=i}^p 
\hat{\omega}_{ij} \left(\frac{y_{n+r} - \bar{y}}{s_y} \right)  \right) \right),
\end{equation}

\section*{Stationarity Conditions}

A stochastic process is said to be strictly stationary when the joint distribution function does not change with a shift in the time \cite{bookwei2006,boxjen76}. However, this definition is too restrictive and can be relaxed by the concept of weak stationarity. A weak stationary stochastic process has expected value constant across time, the second moment finite and its autocovariance is a function of the $|t_1 -  t_2|$.  Stationary is a very important properties because a lot of results such as the law of large numbers and central limit theorem still are holds for stationary process. Furthermore, the sample means is only a good predictor of future behavior if it were constant.

Assuming some assumptions hold valid, results to elucidate the stationarity condition of the GARNN models will be presented.

\textbf{Remark 1} Suppose that a GARNN model has identity link function and a bounded activation function. If $\textbf{x}_t^{'}\boldsymbol{\beta} = \beta_0 \phantom{1} \forall \phantom{1} t$ then process $\left\{y_t\right\}$ is asymptotically stationary.
 
Considering $y_t = \mu_t + a_t$, where $a_t$ is white noise (Benjamin et. al. 2003). Under conditions of Lemma 1, the GARNN  becomes to an autoregressive neural network process (ARNN process) described by \cite{paperTrapletti1970}, who has already proven that ARNN process is asymptotically stationary.

\section*{Ilustrative example}

The data used here was originally modeled by \cite{paperZegerandQaqish1988}, afterward by \cite{paperBenjamin2003} and others. The dataset consisted of the 168 monthly observations of poliomyelitis cases, recorded from the years 1970 to 1983 by the U.S. Centers for Disease Control. The scientific hypothesis is whether the incidence of polio has been decreasing since 1970.  Additionally, \cite{paperBenjamin2003} reported evidence of the presence of the annual and semiannual cycles in the data. I developed R script for carrying out all computations \cite{Rsoftware2018}, i.e., estimations, model selection criteria, analysis of deviance, predictions. Figures were drawn using functions from the ggplot2 library. \cite{bookWickham2012}.

Here, Poisson and negative binomial GARNN models with logarithmic link function and hyperbolic tangent activation function were fitted to the data. The same ad hoc procedure of the model selection was adopted for both models. This procedure for built in two stages. In the first stage, a maximal model of the linear predictor $\mathbf{x}^{'}\boldsymbol{\beta}$ is defined. Number of nodes and autoregressive parameters are selected via AIC. For negative Binomial GARNN model, the $k$ parameter was chosen in the stage as well. Table 1 and 3 displays all  configurations investigated. In the second stage, the paramteres of linear predictor $\mathbf{x}^{'}\boldsymbol{\beta}$ are chosen by analysis of deviance.

For this particular example, the linear predictor of the maximal model is given by:

\begin{equation} \nonumber
\log(\mu_t) = \beta_0 + \beta_1 x_1 + \beta_2 x_2 + \beta_3 x_3 + \beta_4 x_4 + \beta_5 x_5.
\end{equation}
where  $x_1= \cos\left(\frac{2\pi \tilde{t}}{12}\right)$ and $x_2 = \sin\left(\frac{2\pi \tilde{t}}{12}\right)$ are annual seasonality effects; $x_3 = \cos\left(\frac{2\pi \tilde{t}}{6}\right)$ and $x_4 = \sin\left(\frac{2\pi \tilde{t}}{6}\right)$ are semiannual seasonality effects and $x_5 = \tilde{t}$ is the trend effect. The maximal model was based on \cite{paperZegger1988}.

Others nested model are respectively: only constant, only annual cycle, annual and semiannual cycle and maximal model.

\begin{itemize}
\item[-] $\log(\mu_t) = \beta_0$, 
\item[-] $\log(\mu_t) = \beta_0 + \beta_1 x_1 + \beta_2 x_2$, 
\item[-] $\log(\mu_t) = \beta_0 + \beta_1 x_1 + \beta_2 x_2 + \beta_3 x_3 + \beta_4 x_4.$
\end{itemize}

\begin{table}[!ht]
\begin{adjustwidth}{-0.0in}{0in} % comment out/remove adjustwidth environment if table fits in text column.
\centering
\caption{ AIC values for several Poisson GARNN Models}
\begin{tabular}{cccc}
\hline
models & AR & Number Nodes & AIC \\ \hline
1 & 1 & 5 & 573.5615 \\ 
2 & 2 & 5 & 615.7915 \\ 
3 & 3 & 5 & 576.0242 \\ 
4 & 1 & 6 & 556.9602 \\ 
5 & 2 & 6 & 587.3483 \\ 
6 & 3 & 6 & 588.3320 \\
7 & 1 & 7 & 562.8718 \\ 
8 & 2 & 7 & 631.3847 \\ 
9 & 3 & 7 & 755.4314 \\ 
\hline
\end{tabular}
\label{tab1}
\end{adjustwidth}
\end{table}

\begin{table}[!ht]
\begin{adjustwidth}{-0.0in}{0in} % comment out/remove adjustwidth environment if table fits in text column.
\centering
\caption{ AIC values for several Negative Binomial GARNN Models}
\begin{tabular}{ccccc}
\hline
models & AR & Number Nodes & Dispersion & AIC \\ \hline
1 & 1 & 5 & 0.75 & 548.8444 \\ 
2 & 2 & 5 & 0.75 & 560.4201 \\ 
3 & 1 & 10 & 0.75 & 569.9718 \\ 
4 & 2 & 10 & 0.75 & 606.9678 \\
5 & 1 & 5  & 1.5  & 530.9892 \\
6 & 2 & 5  & 1.50 & 545.7424 \\
7 & 1 & 10 & 1.50 & 553.7524 \\
8 & 2 & 10 & 1.50 & 636.2346 \\
\hline
\end{tabular}
\label{tab2}
\end{adjustwidth}
\end{table}

Table \ref{tab1} and \ref{tab2} display AIC values for different Poisson and Negative Binomial models, where for all of them, the linear predictor of the systematic component is the maximal model. Based on AIC, the optimal number of lags and hidden nodes was 1 and 6 for the Poisson GARNN model, whereas for Negative Binomial GARNN was 1 and 5. Moreover, the chosen value of the dispersion parameter was 1.5. The first order autoregressive structure correlation also was considered by \cite{paperZegger1988, paperChan1995, paperDavis2000, paperDavid2009}.

\begin{table}[!ht]
\begin{adjustwidth}{-0.0in}{0in} % comment out/remove adjustwidth environment if table fits in text column.
\centering
\caption{ Analysis of Deviance for Poisson and Negative Binomial GARNN Models}
\begin{tabular}{cccc}
\hline
Distribution & Model & Deviance & P-values \\ \hline
Poisson      & A versus B & 10.51 & 0.00523 \\
Poisson      & B versus C & 17.96 & 0.00013 \\
Poisson      & C versus D & 5.68 & 0.05842 \\
Negative Binomial      & A versus B & 5.16 & 0.07591 \\
Negative Binomial        & B versus C & 7.25 & 0.02656 \\
Negative Binomial        & C versus D & 3.65& 0.16053 \\
\hline
\end{tabular}
\label{tab3}
\end{adjustwidth}
A:  $\log(\mu_t) = \beta_0$; B: $\log(\mu_t) = \beta_0 + \beta_1 x_1 + \beta_2 x_2$, C: $\log(\mu_t) = \beta_0 + \beta_1 x_1 + \beta_2 x_2 + \beta_3 x_3 + \beta_4 x_4.$, D: $\log(\mu_t) = \beta_0 + \beta_1 x_1 + \beta_2 x_2 + \beta_3 x_3 + \beta_4 x_4 + \beta_5 x_5$.
\end{table}

Results from the analysis of deviance revealed no statistical evidence to support the hypothesis that the trend effect is significantly different than zero. For the Poisson GARNN model, annual and semiannual cycles effects were significant (Table \ref{tab3}). These results agree with obtained by several authors that used different models for this same dataset. \cite{paperDavis2000} fitted an autocorrelated latent Poisson process to the data. The authors claimed that the use of a suitable standard error estimator leads to the conclusion that the trend is not statistically significant.  \cite{paperDavid2009} extended the model proposed by \cite{paperDavis2000}, assuming that the random variable follows a negative binomial distribution. Regression coefficients were estimated by maximizing a pseudo-likelihood and were concluded that the negative trend was not significant, using the standard error that includes a latent process.

\begin{figure}[ht] %s state preferences regarding figure placement here
% use to correct figure counter if necessary
%\renewcommand{\thefigure}{2}
\includegraphics[width=\textwidth]{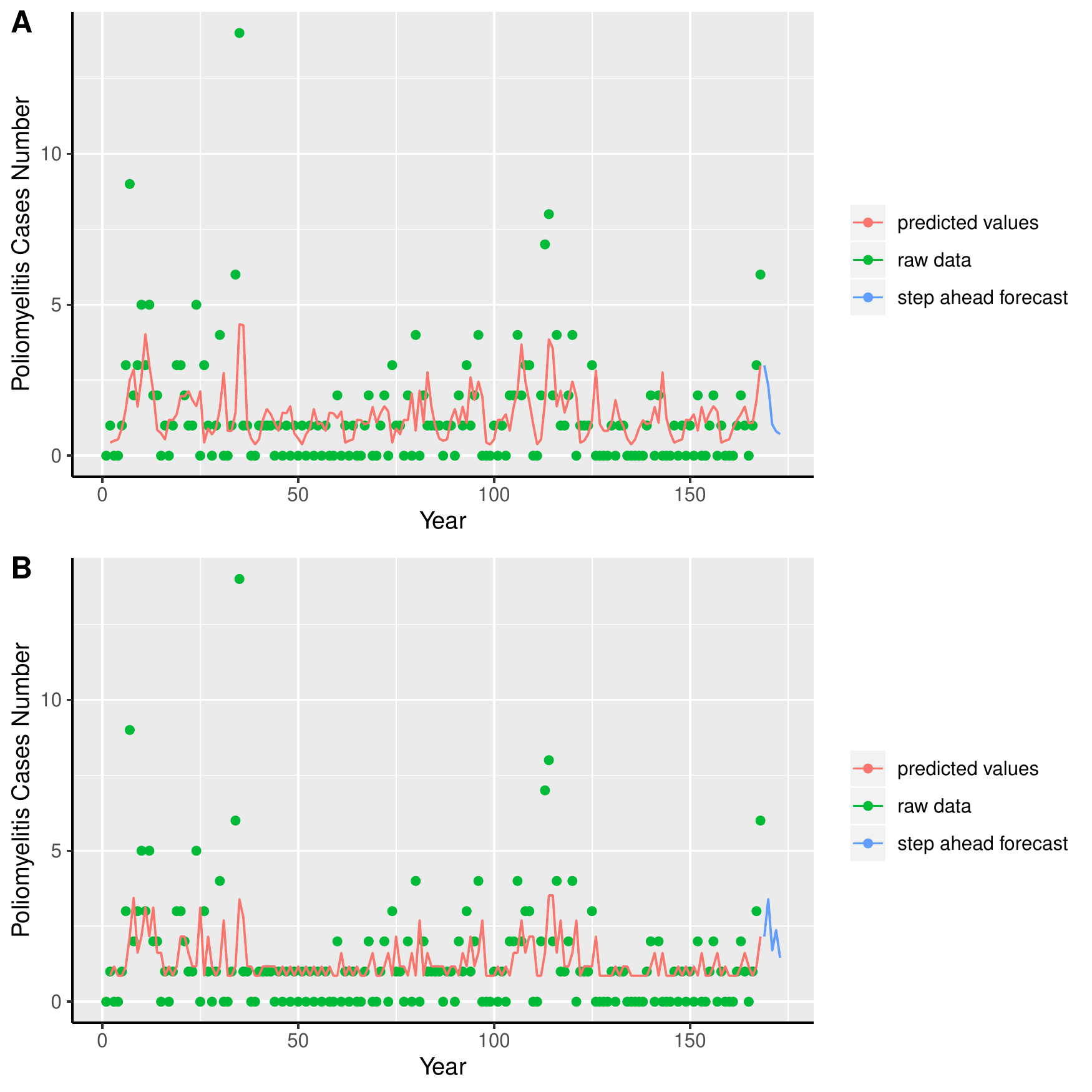}
\caption{\color{Gray} Time series data (green), fitted values (red) and step ahead forecast (blue) for poliomyelitis cases counting. Poisson GARNN (A);  Negative Binomial (B).}
\label{fig1} % \label works only AFTER \caption within figure environment

\end{figure}

On the other hand, for the Negative Binomial GARNN model, none of the explanatory variables were significant. A possible explanation for that is the time series lagged neural network could have captured the seasonality, which becomes the systematic component not so relevant. Comparison between both Poisson and Negative Binomial GARNN models was done. Figure \ref{fig1} shows the fitted and real data for both models. By visual inspection is possible to verify that Negative Binomial fitted slightly better to the data. In the literature, there are some successful cases of fitting neural networks to seasonal time series data. \cite{conferenceCrone2007} fitted multilayer perceptrons in synthetic time series data with different forms of seasonal and trend components. They found that neural networks showed sensitivity to selected architecture decisions but generally provided a robust and reasonable forecasting performance. 

\section*{Discussion}

In this article, a class of GARNN models is proposed as a semiparametric alternative for modeling non-Gaussian time series. I included a time-lagged neural networks in a link function of the generalized linear model. Consequently, the marginal mean depends on the past values of the process. The GARNN model can be considered as a sort of generalized additive model GAM \cite{bookhastie1990} with lagged dependent variables. In the time series context, applications of the GAM and its variants are not new in the literature. For instance, \cite{paperEgondi2012} utilized Poisson GAM for studying the relationship between daily weather and mortality in a population of approximately 60,000 individuals obtained from Nairobi Urban Health and Demographic Surveillance System data during the period 2003-2008. \cite{paperSimpson2018} presented an approach to the estimation of trends in two palaeoenvironmental time series using GAMs. The first dataset was a 150-year bulk organic matter from Small Water, UK. And the second one, a 3,000-year alkenone record from Braya-Sø, Greenland. \cite{paperdeSouza2017} proposed a generalized additive models with principal component analysis to handle with multicolinearity and serial dependence. Subsequently, some properties of this model were discussed theoretically by \cite{paperIspany2018}. However, the novelty of this paper is model the dependency of the observations using a neural network rather than any smooth function. 

In deep learning literature, it is already well known that different loss functions correspond to different assumptions regarding the distribution of the data. \cite{paperDorling2003}. What was done here is unify these distributions in a GLM framework.

The model proposed here is more interpretable and flexible regarding the standard neural network. Adding a neural network to the linear predictor, it is possible interpreting the estimates of the regression coefficients. Additionally, it is still possible taking advantage of the efficiency of the neural network to make predictions. Nowadays, interpretability is very important, because the interpretable machine learning is a hot topic. Although, it could be seen in the illustrative example that sometimes the linear predictor and neural network can compete with each other. 

Still comparing the standard neural networks with the GARNN model, the last one does not need any pre-processing. No require pre-processing is an advantage of the GARNN model regarding standard neural nets. Quite often, the accuracy of the predictions ranges according to kind of the normalization adopted \cite{summarySen2019}.

\bibliography{library}

%This defines the bibliographies style. Search online for a list of available styles.
\bibliographystyle{abbrv}

\end{document}